# A high-gain cladded waveguide amplifier on erbium doped thin-film lithium niobate fabricated using photolithography assisted chemo-mechanical etching

Youting Liang[1, 2], Junxia Zhou[1, 2], Zhaoxiang Liu[2], Haisu Zhang[1, 2, *], Zhiwei Fang[2], Yuan Zhou[3], Difeng Yin[3], Jintian Lin[3], Jianping Yu[3], Rongbo Wu[3], Min Wang[2], and Ya Cheng[1, 2, 3, 4, 5, 6, *]

[1]State Key Laboratory of Precision Spectroscopy, East China Normal University, Shanghai 200062, China.
[2]The Extreme Optoelectromechanics Laboratory (XXL), School of Physics and Electronic Sciences, East China Normal University, Shanghai 200241, China.
[3]State Key Laboratory of High Field Laser Physics and CAS Center for Excellence in Ultra-Intense Laser Science, Shanghai Institute of Optics and Fine Mechanics (SIOM), Chinese Academy of Sciences (CAS), Shanghai 201800, China.
[4]Shanghai Research Center for Quantum Sciences, Shanghai 201315, China.
[5]Collaborative Innovation Center of Extreme Optics, Shanxi University, Taiyuan 030006, China.
[6]Collaborative Innovation Center of Light Manipulations and Applications, Shandong Normal University, Jinan 250358, China.
*Correspondence: Haisu Zhang (hszhang@phy.ecnu.edu.cn), Ya Cheng (ya.cheng@siom.ac.cn).

Erbium doped integrated waveguide amplifier and laser prevail in power consumption, footprint, stability and scalability over the counterparts in bulk materials, underpinning the lightwave communication and large-scale sensing. Subject to the highly confined mode and moderate propagation loss, gain and power scaling in such integrated micro-to-nanoscale devices prove to be more challenging compared to their bulk counterparts. In this work, stimulated by the prevalent success of double-cladding optical fiber in high-gain/power operation, a $Ta_2O_5$ cladding is employed in the erbium doped lithium niobate (LN) waveguide amplifier fabricated on the thin film lithium niobate on insulator (LNOI) wafer by the photolithography assisted chemomechanical etching (PLACE) technique. Above 20 dB small signal internal net gain is achieved at the signal wavelength around 1532 nm in the 10 cm long LNOI amplifier pumped by the diode laser at ~980 nm. Experimental characterizations reveal the advantage of $Ta_2O_5$ cladding in higher optical gain compared with the air-clad amplifier, which is further explained by the theoretical modeling of the LNOI amplifier including the guided mode structures and the steady-state response of erbium ions.

Erbium doped integrated waveguide laser and amplifier, the essential analogs of prevalent erbium doped fiber laser and amplifier, have been intensively investigated under an abundance of material platforms in the recent decades [1-4]. Photonic integration brings various advantages including small footprint, low power consumption and high stability and scalability to the desired waveguide laser and amplifier, though stringent limitations on the attainable gain/power arise due to the highly confined optical mode and moderate propagation loss. In contrary to erbium doped optical fibers which favor effective absorption lengths of tens of meters due to their extremely low propagation loss [5, 6], integrated waveguides feature much higher loss and inhomogeneity limiting the available gain length from few-centimeters to meters. As a result, the erbium doping concentration in integrated waveguide amplifiers ($>10^{20}/cm^3$) is compelled to be much higher than that in fiber amplifiers ($10^{18}$~$10^{19}/cm^3$) to maintain comparable optical gain [7, 8]. High concentration doping instead incurs adverse effects including migration-accelerated energy transfer upconversion (ETU) among excited ions, fast quenching by static ETU among excited ions or clusters, and excited state trapping by host material defects [9, 10]. Such effects degrade the establishment of high population-inversion gain, putting an inherent limit on the attainable gain and engendering delicate optimizations on the doping concentration for specific waveguide configurations.

Integrated lithium niobate (LN) nanophotonics have been broadly investigated in recent years, benefited from the excellent electro-optic, acousto-optic and nonlinear optical properties of lithium niobate and the maturing development of high-quality lithium niobate on insulator (LNOI) wafer and its photonic integration technique [11-14]. Recent demonstration of erbium doped LNOI laser and amplifier has stimulated great interests in developing active LNOI components and passive/active hybrid integration methods [15-24]. The erbium doped LNOI waveguide amplifier holds great promise among such quest due to its broad applicability in optical communication and sensing. Various groups have reported LNOI waveguide amplifiers with diverse footprints and gain performance [20-24], and the maximum small signal internal net gain of ~18 dB is achieved in a 3.6 cm long LNOI amplifier [22].

In this work, an efficient LNOI waveguide amplifier with high erbium doping concentration (~$1.9 \times 10^{20}/cm^3$) is fabricated, featuring above 20 dB small signal internal net gain at the signal wavelength of ~1532 nm pumped by ~980 nm laser diode. The maximum gain is achieved in the $Er^{3+}$: LNOI waveguide with a thin cladding layer of tantalum pentoxide ($Ta_2O_5$). Theoretical modeling based on the mode structures and the steady-state response of erbium ions reveal the role of $Ta_2O_5$ cladding in mitigating the counteracting effect of quenched active ions for high inversion gain. Experimental comparison between the LNOI waveguide amplifiers with/without the $Ta_2O_5$ cladding matches well with the theoretical prediction. Further power scaling of LNOI amplifier by shaping the cladding layer into waveguide structure surrounding the active LN core in analogy to the double-cladding fiber structure which allows efficient cladding pumping is also discussed.

The 600 nm thick $Er^{3+}$-doped Z-cut LN thin film is ion-sliced from the bulk $Er^{3+}$-doped LN crystal by the 'smart-cut' technique, which resides atop of a 2 μm thick silica layer for optical isolation with the underneath 0.5 mm-thick silicon substrate. The LNOI waveguide is fabricated using the photolithography assisted chemo-mechanical etching (PLACE) technique [25, 26], and the fabrication details can be found in our previous works. The schematic for the waveguide cross section is shown in Fig. 1(a), and the LN core features a top width of 2 μm and a bottom width of 6 μm which can be inferred from the top-view reflective microscope image of the fabricated LNOI waveguide shown in Fig. 1(b). A thin layer of $Ta_2O_5$ with ~1 μm thickness is further deposited on top of the LN core. The waveguide input and output facets are prepared by focused ion beam (FIB) milling, with the scanning electron microscope (SEM) image shown in Fig. 1(c). The tilt angle of SEM image is about 52°. The fundamental TE-mode profiles at the pump wavelength (976 nm) and the signal wavelength (1532 nm) employed in the amplifier gain measurement are calculated by finite element method and depicted in Figs. 1(d) and 1(e), for the air-clad and $Ta_2O_5$-clad LNOI waveguides respectively. Clearly, due to the reduced refractive index contrast between LN (refractive index $n_o$~2.212, $n_e$~2.138) and $Ta_2O_5$ (refractive index $n$~2.058), the guided optical modes are penetrated out of the LN core in the $Ta_2O_5$-clad waveguide compared to the air-clad waveguide. The LN core power confinement factors Γ for both waveguides are also calculated and labelled in Figs. 1(d) and 1(e).

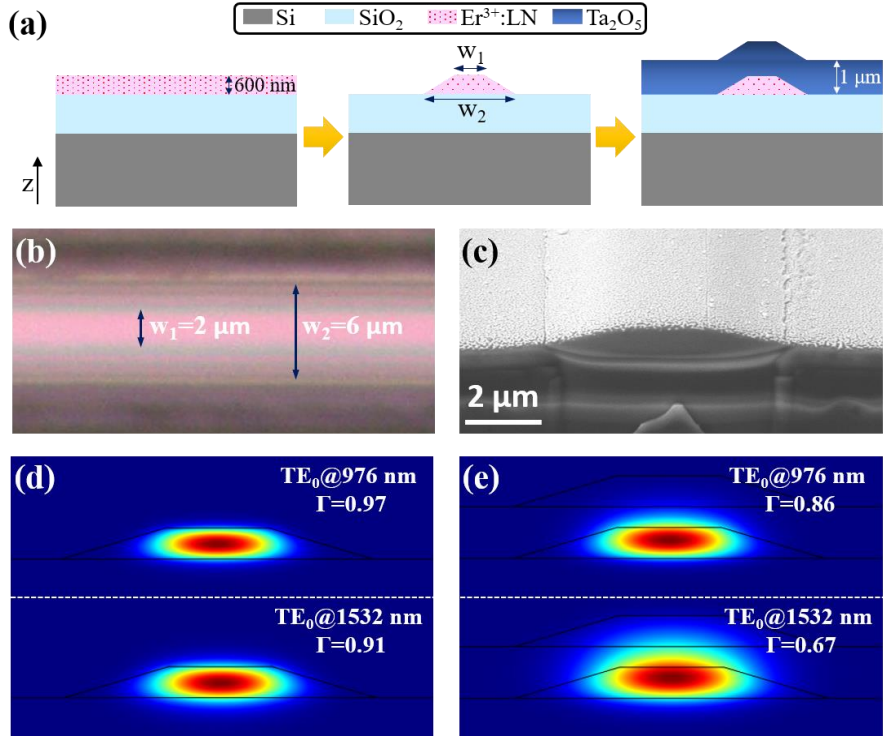

Fig. 1 (a) Cross-sectional schematic of the Z-cut $Er^{3+}$:LNOI wafer, the fabricated LNOI waveguide, and the $Ta_2O_5$-cladding on top of the LNOI waveguide. (b) top view microscope image of the air-clad LNOI waveguide. (c) SEM image of the $Ta_2O_5$-clad LNOI waveguide cross section. (d-e) The simulated fundamental TE modes for the air-clad and $Ta_2O_5$-clad waveguides at the pump and signal wavelengths. The power confinement factor $\Gamma$ is labelled in each panel.

To characterize the gain performance of the LNOI waveguide amplifier, ~10 cm long LNOI waveguides with folded race-track footprint are fabricated and the signal enhancement method is adopted with the experimental schematic shown in Fig. 2. The pump light at 976 nm is provided by a diode laser (CM97-1000-76PM, Wuhan Freelink Opto-electronics), while a continuous-wave C-band tunable laser (TLB 6728, NewFocus) with the wavelength range from 1520 to 1570 nm is used as the signal. The pump and signal waves are combined (separated) by the fiber-based wavelength division multiplexers (WDM) with their respective polarization states being adjusted using in-line fiber polarization controllers at the input (output) port of the integrated amplifier. The combined pump/signal is injected into and coupled out of the amplifier by lensed fibers and bidirectional pumping is employed. The coupling efficiencies of both lensed fibers are estimated to be ~10 dB per facet and all the powers given below are converted into the injected on-chip powers. The output signals are measured by an optical spectrum analyzer (OSA: AQ6370D, YOKOGAWA). A digital camera photograph of the excited $Ta_2O_5$-clad $Er^{3+}$: LNOI waveguide amplifier is also shown in Fig. 2 displaying the clearly visible green upconversion fluorescence from erbium ions along the full waveguide length.

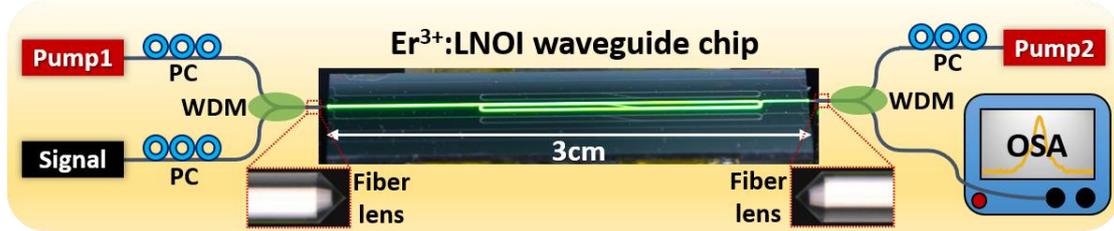

Fig. 2 The experimental setup for optical gain measurement. PC, polarization controller. WDM, wavelength division multiplexer. OSA, optical spectrum analyzer. The digital camera photograph of the excited $Er^{3+}$:LNOI waveguide chip is shown in the center.

The optical gain of the LNOI waveguide amplifiers with/without $Ta_2O_5$ cladding is first measured at an injected signal power of -24 dBm and variable pump powers. The signal wavelength is tuned to 1532 nm corresponding to the peak absorption and emission cross sections of erbium ions [27]. The internal net gain of the amplifier is deduced from the signal enhancement factor as:

$$G = 10\log_{10}\frac{S^w}{S^o} - \alpha L$$

where $S^w$ ($S^o$) is the output signal power with (without) pumps, $\alpha$ is the propagation loss experienced by the signal alone including both the scattering loss as well as the absorption loss, $L$ is the waveguide length.

The measured small signal internal net gain for both waveguides is plotted in Fig. 3, where red circles represent gain values from $Ta_2O_5$-clad waveguide and blue squares denote gain values from air-clad waveguide. The signal propagation loss used to calibrate the internal net gain is estimated from the resonator-loss measurement of the $Er^{3+}$: LNOI whispering-gallery-mode (WGM) microcavity fabricated using the same LNOI wafer without $Ta_2O_5$ cladding. The optical transmission curve around the resonant wavelength of 1532 nm is also depicted in the inset of Fig. 4 which gives an intrinsic quality factor of $Q \approx 4 \times 10^5$. Thus, the propagation loss at $\lambda=1532$ nm is obtained by $\alpha=2\pi n_g/(Q\lambda)$ as 0.98 dB/cm, with the group index $n_g=2.2$ deduced from the free spectral range measurement. It is assumed that the measured propagation loss being dominated by the absorption loss from ground-state $Er^{3+}$ since the passive scattering loss should be <0.1 dB/cm due to the employed fabrication method (photolithography assisted chemical etching) which can reduce the waveguide surface roughness down to subnanometer scale [25]. Then the signal propagation loss in $Ta_2O_5$-clad waveguide is scaled by the power confinement factors in the LN core as $\alpha_{TaO}= \alpha\Gamma_{TaO}/\Gamma_{Air}=0.72$ dB/cm, where $\Gamma_{Air}$ and $\Gamma_{TaO}$ are the power confinement factors for the air-clad and $Ta_2O_5$-clad waveguides, respectively.

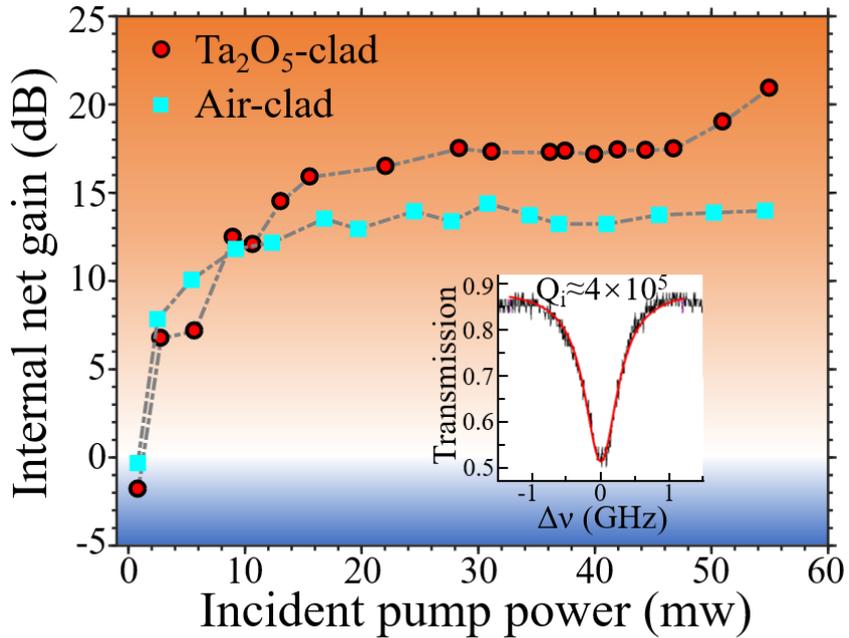

Fig. 3 The internal net gain measured from the air-clad (blue squares) and $Ta_2O_5$-clad (red circles) amplifiers. The optical transmission curve around the resonant wavelength of 1532 nm in the $Er^{3+}$:LNOI WGM-microcavity is shown in the inset.

It can be clearly seen from Fig. 3 that the internal net gain increases rapidly with increasing pump powers and then gradually saturates due to reduced inversion of erbium ions at high gain regime. Notably, the net gain provided by $Ta_2O_5$-clad waveguide amplifier is higher than the air-clad amplifier when the injected pump powers are higher than 10 mW, and the maximum net gain from $Ta_2O_5$-clad amplifier is above 20 dB at the maximum pump power of 55 mW. This observation indicates that the $Ta_2O_5$ cladding can promote the build-up of higher inversion gain along the 10 cm long waveguide amplifier compared to its counterpart without this cladding (air-clad amplifier).

The gain response for the $Ta_2O_5$-clad amplifier is further measured at the pump power of 50 mW and variable signal powers. The result is plotted in Fig. 4. It can be easily noticed that the internal net gain provided by the amplifier decreases with increasing signal powers, arising from the depletion of excited state erbium ions by stimulated emissions induced by high power signals. Besides, the gain bandwidth of the amplifier is also characterized by tuning the signal wavelengths at small signal powers, featuring a broad net gain from 1525 nm to 1565 nm with the maximum gain peaked around 1532 nm. This gain bandwidth matches well with the $Er^{3+}$ fluorescence band and the observed gain bandwidths in $Er^{3+}$: LNOI waveguide amplifiers demonstrated recently [20-24].

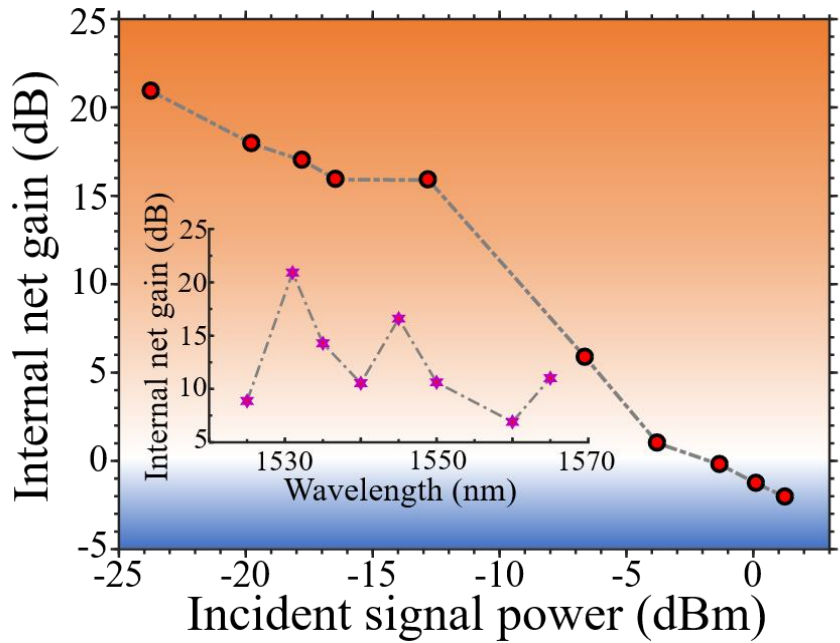

Fig. 4 The internal net gain from $Ta_2O_5$-clad amplifier for increasing input signal powers. The small signal gain values at different signal wavelengths are shown in the inset.

The advantage of the $Ta_2O_5$-cladding in promoting the optical gain of $Er^{3+}$: LNOI waveguide amplifier deserves a deeper investigation. To this end, an established amplifier model based on the optical mode structures and the steady-state response of erbium ions including the migration-accelerated ETU and the fast quenching is employed to simulate the gain behaviors of both the $Ta_2O_5$-clad and air-clad amplifiers [8, 9]. The waveguide length is fixed at 10 cm in the simulation, and the passive propagation loss is set at 0.1 dB/cm. Spectroscopic parameters of $Er^{3+}$ are adopted from the values measured in the $Er^{3+}$-doped bulk LN crystals [27-30]. The optical gain as a function of the input pump and signal powers is first simulated without the contribution of quenched ions for both waveguide amplifiers. The obtained gain values are plotted in pseudo-colors in Figs. 5(a) and 5(b) for the air-clad and $Ta_2O_5$-clad LN waveguides, respectively. The grey dashed lines in Figs. 5(a) and 5(b) denote the '0' dB gain curves delimiting the practical loss region (blue-shaded) and gain region (red-shaded) in the gain maps. Identical behaviors can be observed with comparable gain values for both waveguide amplifiers.

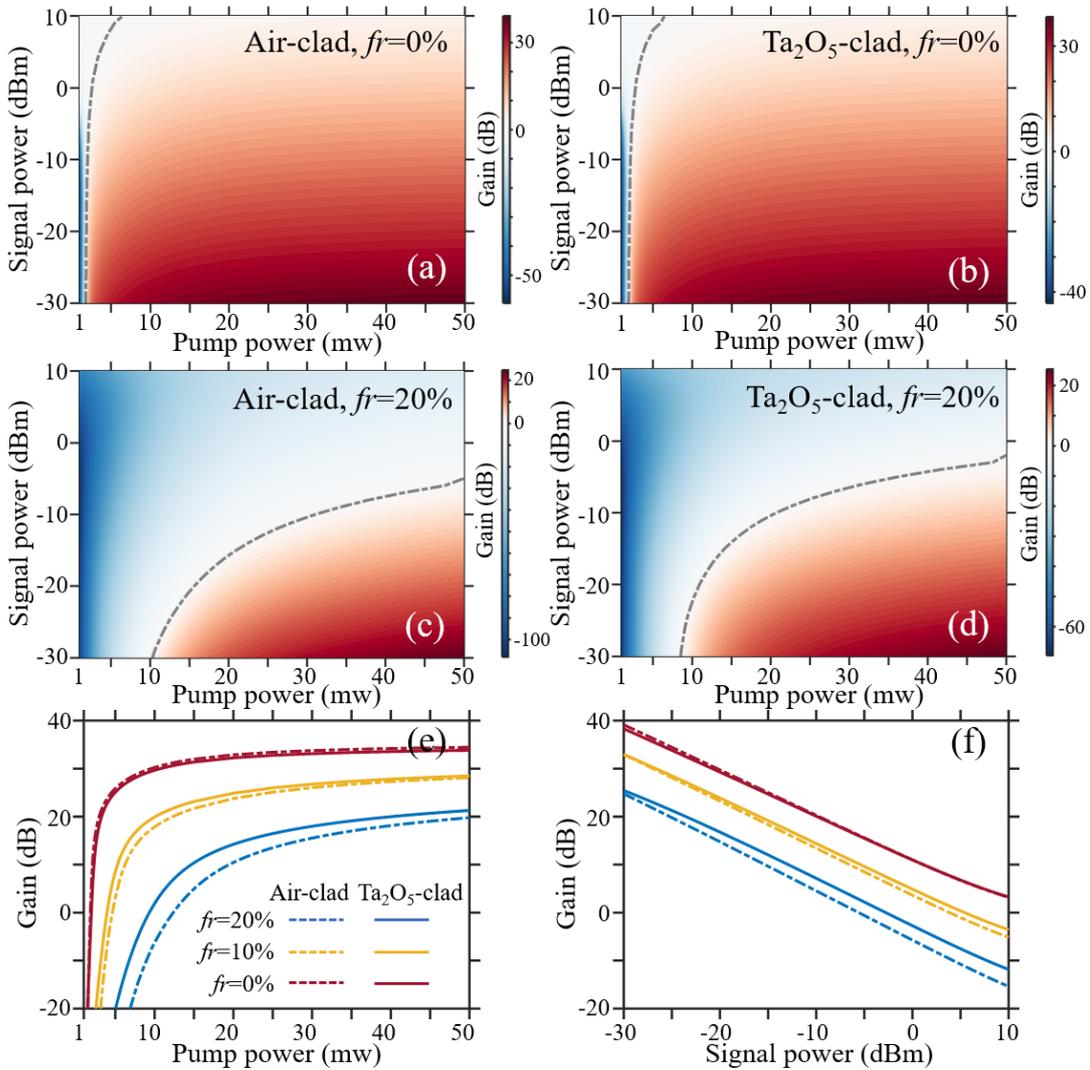

Fig. 5 The simulated gain response for both waveguide amplifiers. The gain map calculated at each pair of signal and pump powers is depicted in pseudo-colors for air-clad amplifier (a) and (c), Ta$_2$O$_5$-clad amplifier (b) and (d). The fractions of quenched erbium ions are labelled in each panel. The gain curve vs pump power and the gain curve vs signal power are plotted in (e) and (f). The same figure legend is applied for (e) and (f).

The gain behaviors of both amplifiers including the contribution from quenched ions are further investigated. The gain values obtained by supposing 20% of active ions are fast quenched are depicted in pseudo-colors in Figs. 5(c) and 5(d), where the grey '0' dB curves delimit the gain and loss regions as well. It can be clearly seen that the presence of quenched ions greatly reduces the optical gain, due to their detrimental absorption of the pump and signal powers without contributing to the inversion gain. Moreover, it can be noticed that both the maximum small signal gain and the area of practical gain region are larger for the Ta$_2$O$_5$-clad amplifier compared to that of the air-clad amplifier. To be more specific, the small signal gain at a fixed signal power (-25 dBm) and increasing pump powers and the signal gain at a fixed pump power (50 mW) and increasing signal powers are plotted in Figs. 5(e) and 5(f) for both waveguide amplifiers with three different fractions of quenched ions $f_r$=0%, 10%, 20%, respectively. It can be easily inferred from the gain curves that, besides the normal gain saturation behaviors due to the depletion of inversion gain by the pump and signal at high powers, the optical gain provided by Ta$_2$O$_5$-clad amplifier is higher than the gain by air-clad amplifier at larger fractions of quenched erbium ions. Since the quenched ions are unavoidable in the Er$^{3+}$: LNOI amplifier with high doping concentration, the simulation results indicate the better gain response of Ta$_2$O$_5$-clad amplifier than the air-clad amplifier in practical conditions.

The revealed gain responses from the simulation results match qualitatively with the observed gain dynamics and superior gain performance of the Ta$_2$O$_5$-clad amplifier compared to the air-clad amplifier, though a quantitative agreement between experiment and simulation necessitates further characterizations of the spectroscopic response of Er$^{3+}$ in the ion-sliced LN thin film and the fraction of quenched ions which is beyond the scope of current work. Nevertheless, the above simulation pinpoints the role of cladding in reducing the counteracting effect by quenched ions, which can be understood by considering the mode structures in both waveguide amplifiers. A moderate fraction of optical power is guided outside the active LN core in the Ta$_2$O$_5$-clad waveguide, which can be immune from the detrimental absorption of pump and signal powers by the quenched ions. Though the guided power in the core is thus lower which decreases the peak intensities and thus the population inversion gain, this reduction is compensated by the more uniform inversion gain along the full waveguide length due to the suppression of quenched absorption at high saturating powers for the pump and signal. Meanwhile, additional simulations have been done for the maximum output signal powers provided by the amplifiers at each pump and signal powers. The saturated power provided by the Ta$_2$O$_5$-clad amplifier is also higher than the air-clad amplifier for certain quenched factions.

Further power scaling of the Er$^{3+}$: LNOI waveguide amplifier can be conceived by optimizing the cladding materials and guiding geometries. Important clues can be borrowed from the prevalent double-cladding optical fiber design, which allows guiding a large amount of optical power within the inner cladding area which surrounds the much narrow active core [5, 6]. The pump power is progressively absorbed by successive crossing the active core which provides high inversion gain for the signal mode mainly guided in the core. Shaping the cladding material into waveguiding geometry will allow direct cladding pumping which could also pave a way for high power output in the Er$^{3+}$: LNOI waveguide amplifiers.

In conclusion, above 20 dB small signal internal net gain in a Ta$_2$O$_5$-clad Er$^{3+}$: LNOI waveguide amplifier has been demonstrated. Experimental characterizations reveal the advantage of Ta$_2$O$_5$ cladding in gain performance compared with the Er$^{3+}$: LNOI amplifier without cladding. Theoretical modelling pinpoints the role of cladding in mitigating the quenched ion absorption and providing better gain performance for the Er$^{3+}$: LNOI waveguide amplifier. The demonstrated high-gain Er$^{3+}$: LNOI amplifier hold great promise in a broad spectrum of applications such as optical communication, integrated microcomb and precision metrology.

## Acknowledgements

The authors thank Dr. Yang Liu and Prof. Wenxue Li from East China Normal University for their technical support and helpful discussions.